\documentclass[12pt]{article}

\usepackage{graphicx}

\newcommand{\be}{\begin{equation}}
\newcommand{\ee}{\end{equation}}

\newcommand{\eps}{\epsilon}

\begin{document}

\begin{flushright}
	    \  %%% Version 1.3
\end{flushright}

\begin{center}

\vskip 0.5cm

{\Large {\bf Full-Vector Analysis of a Realistic Photonic Crystal Fiber}}\\

\vskip 1cm

{A. Ferrando$^1$, E. Silvestre$^1$, J. J. Miret$^1$,
 M. V. Andr\'es$^2$, and P. Andr\'es$^1$}

\vskip 0.5cm

{$^1$ Departament d'\`{O}ptica, Universitat de Val\`encia.}\\[-0.4cm]
{E-46100 Burjassot (Val\`{e}ncia), Spain.}\\[-0.4cm]
{$^2$ Institut de Ci\`encia dels Materials, Universitat de Val\`encia.}\\[-0.4cm]
{E-46100 Burjassot (Val\`{e}ncia), Spain.}\\

\vspace{1cm}

{\bf Abstract}

\begin{quotation}
\small We analyze the guiding problem in a realistic photonic crystal fiber using a novel full-vector modal technique, a biorthogonal modal method based on the nonselfadjoint character of the electromagnetic propagation in a fiber. 
Dispersion curves of guided modes for different fiber structural parameters are calculated along with the 2D transverse intensity distribution of the fundamental mode.
Our results match those achieved in recent experiments, where the feasibility of this type of fiber was shown.
\end{quotation}

\vspace{1cm}

{\small
\begin{tabular}{rl} OCIS codes: & 060.2270 fiber characterization, \\[-0.4cm]
		& 060.2280 fiber design and fabrication, \\[-0.4cm]
		& 060.2430 fibers, single mode.
\end{tabular}
}

\end{center}

\clearpage

Periodic dielectric structures (photonic crystals) have raised a growing interest in the last years because they exhibit very interesting optical features. The most relevant property of a photonic crystal is the possibility of generating photonic band gaps for certain geometries \cite{yablonovitch}. This effect has been observed in both 2D and 3D structures in the form of absence of light propagation for a specific set of frequencies
(see Ref.~\cite{2d-3dphotonic} and references therein). A related phenomenon occurring in photonic crystal structures is light localization at defects.

Although the previous phenomena were first observed and analyzed 
in bulk structures, there exists a connected effect of potential 
interest for light propagation in dielectric crystals which have 
a 2D periodicity in the $x$-$y$ plane broken by the presence of a 
defect, but are continuous in the $z$ direction. 
The physical realization of such a structure is what 
is called a photonic crystal fiber. This is a thin silica 
fiber having a regular structure of holes that extend themselves 
along the whole fiber length. If one of these 
holes is absent, the transverse dielectric periodicity gets broken 
and a defect appears. The fact that light may be trapped at defects 
turns here into a propagation feature. Consequently, the bound states of the 2D transverse problem 
(2D trapped states of light) become the guided modes of the fiber propagation problem.
The experimental feasibility of these fibers has been proven recently \cite{bath}. A preliminary interpretation of their behavior involving the concept of effective refractive index is presented in Ref.~\cite{bath2}.
A robust single-mode structure was observed for an unusually wide range of wavelengths, a very remarkable property not present in ordinary fibers. 

Our interest lies in giving an appropriate treatment of the realistic problem of a photonic crystal fiber by modeling and solving efficiently its transverse 2D structure. We next proceed to present an approach in which the full-vector character of light propagation in fibers is taken into account. It is an adapted version of our biorthonormal-basis modal method \cite{enrique}. In this way, a realistic 2D periodic structure with a central defect is properly implemented allowing us to analyze different fiber designs. As we will see, our results agree with those experimentally measured
and, at the same time, predict different interesting behaviors for some particular designs.

Guided modes in an inhomogeneous fiber verify a set of dimensionally reduced equations 
involving the transverse coordinates $x$ and $y$ exclusively \cite{snyder}.
This set of equations is obtained from Maxwell's 
equations by assuming the electromagnetic field to be 
monochromatic in time and to have a harmonic 
dependence on $z$ (i.e., the field has a well-defined propagation constant $\beta$). 
In terms of the transverse components of the magnetic and electric field, $h_t=\left(\begin{array}{c} h_x \\ h_y \end{array} \right)$ and $e_t=\left(\begin{array}{c} e_x \\ e_y \end{array} \right)$, 
these equations can be rewritten as \cite{enrique}
\begin{equation}
	L h_t = \beta^2 h_t, \hspace{1cm} 
	L^{\dagger} \bar{e}_t = \beta^{\ast 2} \bar{e}_t
	\label{2deq}
\end{equation}
where $\bar{e}_t =\left(\begin{array}{c} e_y^\ast \\ - e_x^\ast \end{array} \right)$, 
$L^\dagger$ is the adjoint operator of $L$, $^\ast$ denotes the complex conjugate 
operation, and each element $L_{\rho\sigma}$ of the matrix differential operator $L$ 
has the form 
\begin{equation}
	L_{\rho\sigma} \equiv (\nabla^2 + k^2 n^2 )  \delta_{\rho\sigma}
	- (\eps_{\rho\tau}{\frac{\nabla_\tau n^2}{n^2}})
	  (\eps_{\sigma\upsilon}{\nabla_\upsilon}), \hspace{1cm}
	\rho,\sigma,\tau,\upsilon = x,y,
\end{equation}
where $\eps_{\rho\sigma}$ is the completely antisymmetric tensor in 2D, $n$ 
the refractive index, and $k$ the free-space wave number. Of course, $\nabla^2$ 
is the Laplacian operator and $\nabla_\sigma$ the transverse components of 
the gradient operator. 
Let us notice that the general problem of light propagation in a fiber, even for 
nonabsorbing materials (when $n^2$ is real), involves the nonhermitian operator $L$. 

The most relevant property of Eq.~(\ref{2deq}) is that it constitutes a system of eigenvalue equations for the $L$ operator and its adjoint $L^\dagger$ (something that it is far from obvious when one starts from the reduced equations written in terms of $h_t$ and $e_t$ ---instead of $\bar{e}_t$---, see, for instance, Ref.~\cite{snyder}). This feature is crucial in our approach to the full-vector problem. Because $h_t$ and $\bar{e}_t$ are the eigenfunctions of the $L$ and $L^\dagger$ operators, respectively, they are closely related. In fact, they verify what it is called the biorthogonality relation, $\langle \bar{e}_t^n |h_t^m \rangle=\delta_{nm}$ \cite{morse}. The biorthonormality property of the $\{ L,L^\dagger \}$ eigenfunctions, $\{ h_t^m , \bar{e}_t^n \}$, guarantees the expansion of any squared integrable function in terms of either the $L$ or the $L^\dagger$ eigenfunctions, in complete analogy with the hermitian case.
For the same reason, matrix elements can always be defined in terms of the ``biorthogonal modes" of an arbitrary $\{L,L^\dagger \}$ system. 

The main goal of our approach is to transform the problem of solving the system of differential equations (\ref{2deq}) (including highly non trivial boundary conditions sometimes) into an algebraic problem involving the diagonalization of the $L$-matrix. The spectrum of the $L$-matrix will be formed in general by 2D bound states and continuum states. In terms of fiber propagation, the bound states of the $L$ spectrum are guided modes since, despite the finite width of the fiber, the fields exhibit a strong decay in the transverse direction. On the contrary, states from the continuum radiate radially and thus are not guided by the fiber.

The choice of an appropriate auxiliary basis is very important for an efficient implementation of our method. In the particular case of a photonic crystal fiber this election must be specially accurate. The main reason is that the complicated spatial structure of the refractive index in a realistic case can transform the actual computation of the $L$-matrix elements into an impossible task. Realistic simulations must contemplate as many as nearly one hundred 2D step-index individual structures (the air holes of the photonic crystal fiber).
Therefore, a {\em brute force} computation of the matrix elements becomes useless in practice due to critical loss of precision. 

The implementation of the dielectric structure is carried out by putting the system in a finite 2D box (of dimensions $D_x$ and $D_y$) and requiring the fields to fulfill periodic boundary conditions in the $x$ and $y$ directions. So, we create an artificial lattice by replicating the original almost periodic structure, including the central defect, in both transverse directions. This new superlattice is made of copies of the original cell covering the whole two dimensional transverse plane. Although the original cell is not periodic, the whole superlattice really is. The periodicity requirement implies we can expand the 2D electromagnetic fields in a discrete Fourier series in terms of plane waves determined by the exponential functions 
$f_{\vec{n}}(\vec{x}_t)=\exp(i \vec{k}_{\vec{n}} \cdot \vec{x}_t)$, where $\vec{k}_{\vec{n}}= 2 \pi (\frac{n_x}{D_x},\frac{n_y}{D_y})$ is the discretized transverse wave vector. This is equivalent to say we are choosing the $\{ f_{\vec{n}}(\vec{x}_t) \}$ set as the auxiliary basis of our Hilbert space necessary to define the matrix elements of the $L$-operator (note that the $f_{\vec{n}}$ functions are orthogonal). 

A crucial property of the above plane-wave basis is that,
due to periodicity of the superlattice and despite it is defined in a finite volume
---the unit cell of the superlattice of size $D_x$ times $D_y$---, it is translationally invariant. 
The presence of this symmetry shown by our auxiliary basis turns out critical for the feasibility of the method. The advantage of the translation symmetry is twofold. On the one hand, it allows us to relate easily any matrix element of the operator representing a hole at an arbitrary position with that representing a hole at the origin of coordinates. 
Since the whole matrix of the photonic crystal fiber structure can be written as a sum over all matrices representing each one of the substructures (holes), and since these substructures are identical (although differently located), it is possible to reduce the problem to the calculation of one single matrix.
On the other hand, the calculation of any element of this single matrix can be worked out analytically in this basis (we assume a circular step-index profile for the hole). On top of that, and because of the symmetry properties of a realistic hexagonally-centered configuration of holes, the sum over the set of points where the holes are located can also be analytically solved. Consequently, the choice of periodic plane waves of the superlattice as a basis to define the matrix elements of the realistic photonic fiber operator $L$ leads to a crucial simplification. The problem of critical loss of precision due to the complex spatial structure of the photonic crystal fiber is, in this way, overcome.

We have simulated a realistic photonic crystal fiber characterized by a hexagonal distribution of air holes with a central defect. The hole radius $a$, the horizontal distance between the center of two consecutive holes ---or pitch--- $\Lambda$,
and the wavelength of light $\lambda$ are free parameters that we have changed at will. The height of the refractive index step is also free, although we have kept it constant for comparison purposes. We have simulated first a realistic air-filled fiber with parameters $a= 0.3\;\mu$m and $\Lambda=2.3\;\mu$m. We have focussed on this particular design because the intensity distribution for the guided mode in this structure has been measured experimentally for a wavelength of $\lambda=632.8\;$nm. Experimental measures also show that the guided mode in this fiber remains single in a remarkably wide wavelength range, extending from 337~nm to 1550~nm
\cite{bath}. Our simulation allows us to evaluate the eigenvalues of the $L$-operator at any wavelength and thus to calculate the modal dispersion curves for the fiber under consideration in an even wider range of wavelengths (see Fig.~1). The single-mode structure is formed by a polarization doublet. Our results completely agree with the previous experimental results as they account for the existence of a robust single-mode structure nearly at all wavelengths for the above fiber parameters. We include in Fig.~1 the envelope of the radiation modes which is refered to as the cladding index (i.e. the effective refractive index of the photonic crystal).

Since the diagonalization procedure of the full-vector operator $L$ generates not only the set of eigenvalues but also their respective eigenvectors, we can also evaluate the transverse intensity distribution of the electromagnetic field for the guided mode. The result for one of the polarizations is shown in Fig.~2 for $\lambda=632.8$~nm and reproduces, with excellent accuracy, that experimentally measured \cite{bath}. We have also calculated the transverse intensity of the guided mode at very different wavelengths and for both polarizations. In all cases, the intensity profile is very similar to that shown in Fig.~2. In this way, we have checked the robust character of the single mode structure under changes in the wavelenght of light. This fact agrees with the behavior of the dispersion curves mentioned previously.

Besides simulating this remarkable structure, we have also simulated a number of different fiber designs by changing the pitch $\Lambda$ and the hole radius $a$. By analyzing the dispersion curves of these differently fibers, we have found a richer modal structure in some of them. In the example shown in Fig.~3 there exist, besides the fundamental doublet, two other polarization doublets. Unlike conventional fibers, the number of modes does not increase with the light wave number $k$. The number of guided modes gets stabilized above a $k$-threshold, or equivalently it remains constant for wavelengths smaller than a threshold wavelength. For particular designs one can get guiding structures in which this constant number is just one. In such a case one obtains and ``endlessly'' single-mode fiber as the one reported in Ref.~\cite{bath}.
This is a very unconventional property shown by photonic crystal fibers.

In a conventional fiber, the cladding refractive index is nearly constant and then its V-value, the optical ``volume" (or phase space) of the fiber, grows with $k$. This fact permits to accommodate an increasing number of guided modes inside the fiber as the wavelength is reduced. In a photonic crystal fiber the periodic structure responsible for light trapping at the central defect creates a dependence on the effective refractive index of the cladding such that a much more weakly $k$-dependent V-value is generated. The optical ``volume" becomes then practically independent of the wavelength for large values of $k$ and, consequently, so do the number of guided modes.

\vspace{\baselineskip}

Financial support from the Generalitat Valenciana (grant GV96-D-CN-05-141) is acknowledged.

\clearpage

\section*{Figure captions}

\begin{itemize}

\item Figure~1. Modal dispersion curves extending from $\lambda=300$~nm to 
$\lambda=1600$~nm for a single-mode photonic crystal-fiber structure with 
$a=0.3\;\mu$m and $\Lambda=2.3\;\mu$m. In this plot, the variation of the 
mode index for both polarizations coalesce in a single curve.

\item Figure~2. Transverse intensity distribution for the $x$-polarized 
guided mode of the photonic crystal fiber described in Fig.~1 for 
$\lambda=632.8\;$nm.

\item Figure~3. Same as in Fig.~1 but with $a=0.6\;\mu$m. Here, the two higher order polarization doublets are slightly shifted each other.

\end{itemize}

%%% \noindent
%%% \includegraphics{c:/user/mora/cristal/fcf/1998-ol/f1.eps}

%%% \noindent
%%% \includegraphics{c:/user/mora/cristal/fcf/1998-ol/f2.eps}

%%% \noindent
%%% \includegraphics{c:/user/mora/cristal/fcf/1998-ol/f3.eps}


\clearpage

\begin{thebibliography}{99}
\bibitem{yablonovitch}
E. Yablonovitch, J. Opt. Soc. Am. B {\bf 10}, 283 (1993).

\bibitem{2d-3dphotonic}
P. St. J. Russell, T. A. Birks, and F. D. Lloyd-Lucas, 
in {\em Confined Electrons and Photons}, E. Burstein and C. Weisbuch, eds. 
(Plenum Press, New York, 1995), p. 585.

\bibitem{bath}
J. C. Knight, T. A. Birks, P. St. J. Russell, and D. M. Atkin, 
Opt. Lett. {\bf 21}, 1547 (1996); Opt. Lett. {\bf 22}, 484 (1997).

\bibitem{bath2}
T. A. Birks, J. C. Knight, and P. St. J. Russell, Opt. Lett. {\bf 22}, 961 (1997).

\bibitem{enrique}
E. Silvestre, M. V. Andr\'es, and P. Andr\'es,
J. Lightwave Technol. {\bf 16}, 923 (1998).

\bibitem{snyder}
A. W. Snyder and J. D. Love, {\em Optical Waveguide Theory} 
(Chapman and Hall, London, 1983), pp. 595-606.

\bibitem{morse}
P. M. Morse and H. Feshbach, {\em Methods of Theoretical Physics}, part I 
(McGraw-Hill, New York, 1953), pp. 884-886.

\end{thebibliography}
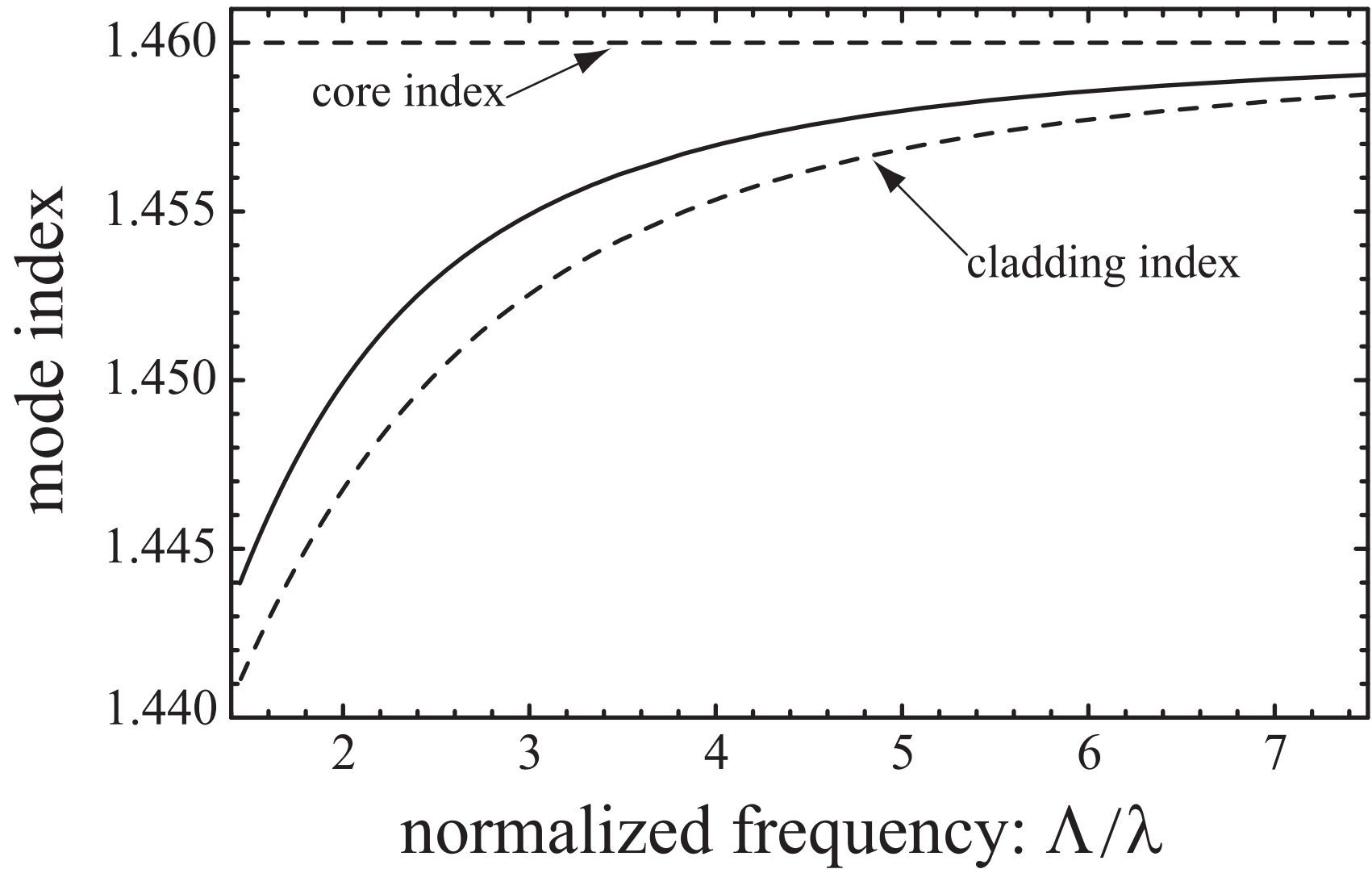
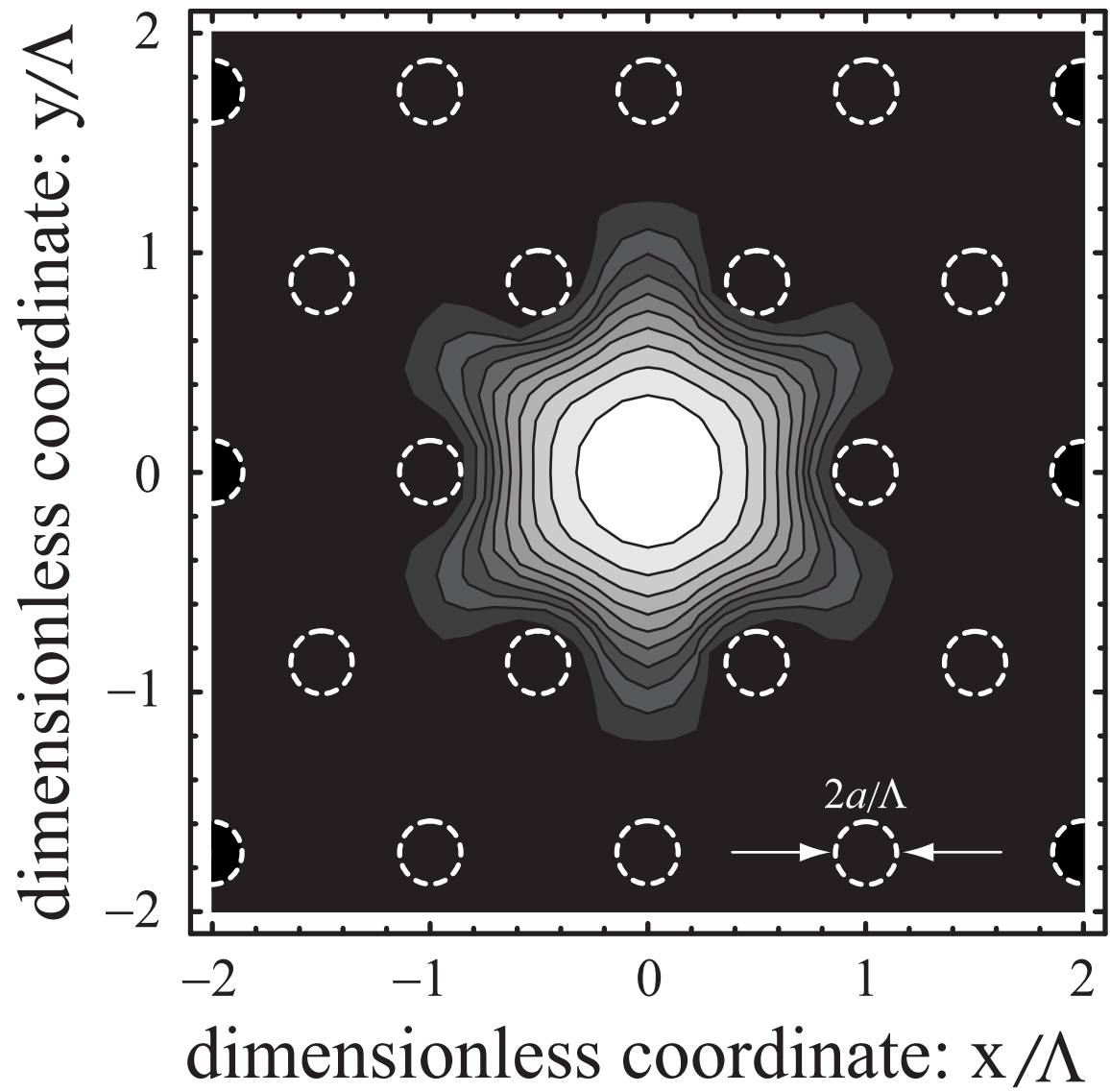
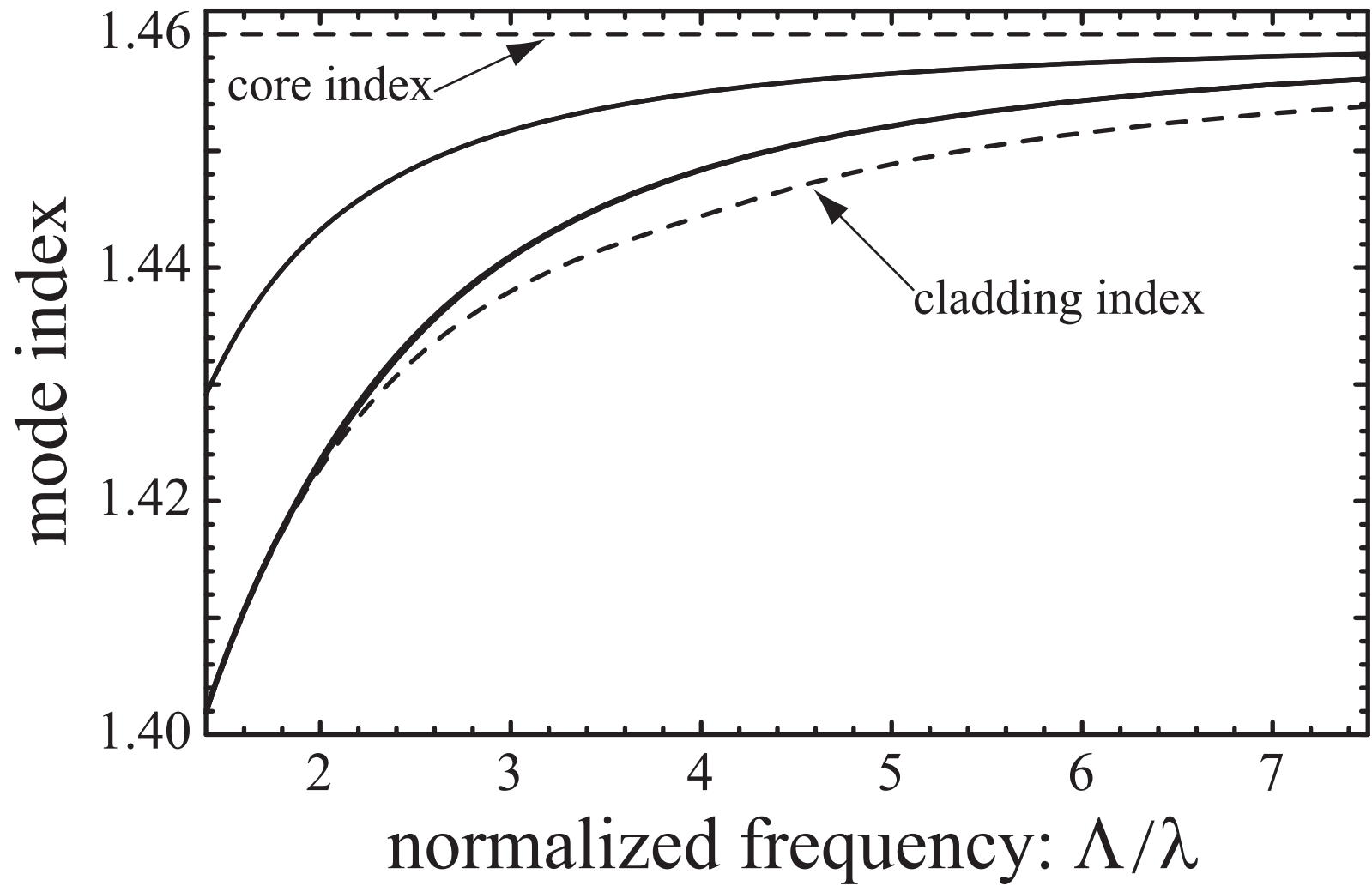
\end{document}